\documentstyle[psfig,multicol,osa]{revtex}
\begin{document}

\setlength{\textwidth}{180mm}
\setlength{\textheight}{240mm}
\setlength{\parskip}{2mm}

\title{Nonlinear transmission and light localization in photonic
crystal waveguides}
\author{Sergei F. Mingaleev and Yuri S. Kivshar}

\address{Nonlinear Physics Group, Research School of Physical
Sciences and Engineering \\
Australian National University, Canberra ACT 0200, Australia}

\maketitle

\begin{abstract}
We study the light transmission in two-dimensional photonic
crystal waveguides with embedded nonlinear defects. First, we
derive the effective discrete equations with long-range
interaction for describing the waveguide modes, and demonstrate
that they provide a highly accurate generalization of the familiar
tight-binding models which are employed, e.g.,  for the study of
the coupled-resonator optical waveguides. Using these equations,
we investigate the properties of straight waveguides and waveguide
bends with embedded nonlinear defects and demonstrate the
possibility of the nonlinearity-induced bistable transmission.
Additionally, we study localized modes in the waveguide bends and
(linear and nonlinear) transmission of the bent waveguides and
emphasize the role of evanescent modes in these phenomena.
\end{abstract}

\begin{multicols}{2}

\narrowtext


\section{Introduction}

Photonic crystals are usually viewed as an optical analog of
semiconductors that modify the properties of light similar to a
microscopic atomic lattice that creates a semiconductor band-gap
for electrons \cite{book}. One of the most promising applications
of photonic crystals is a possibility to create compact integrated
optical devices \cite{book_sakoda}, which would be analogous to
the integrated circuits in electronics, but operating entirely
with light. Replacing relatively slow electrons with photons as
the carriers of information, the speed and bandwidth of advanced
communication systems can be dramatically increased, thus
revolutionizing the telecommunication industry.

To employ the high-tech potential of photonic crystals for optical
applications and all-optical switching and waveguiding
technologies, it is crucially important to achieve a dynamical
tunability of their properties. For this purpose, several
approaches have been suggested (see, e.g., Ref. \onlinecite{john2}). One
of the most promising concepts is based on the idea to employ the
properties of {\em nonlinear photonic crystals}, i.e. photonic
crystals made from dielectric materials whose refractive index
depends on the light intensity. Exploration of nonlinear
properties of photonic band-gap materials is an important
direction of current research that opens a broad range of novel
applications of photonic crystals for all-optical signal
processing and switching, allowing an effective way to create
highly tunable band-gap structures operating entirely with light.

One of the important concepts in the physics of photonic crystals
is related to the field localization on defects. In the
solid-state physics, the idea of localization is associated with
disorder that breaks the translational invariance of a crystal
lattice and supports spatially localized modes with the frequency
outside the phonon bands. The similar concept is well-known in the
physics of photonic crystals where an isolated defect (a region
with different refractive index which breaks periodicity) is known
to support a localized defect mode. An array of such defects
creates a waveguide that allows a directed light transmission for
the frequencies inside the band gap. Since the frequencies of the
defect modes created by {\em nonlinear defects} depend on the
electric field intensity, such modes can be useful to control the
light transmission. From the viewpoint of possible practical
applications, spatially localized states in optics can be
associated with different types of all-optical switching devices
where light manipulates and controls light itself due to the
varying input intensity.

Nonlinear photonic crystals and photonic crystals with embedded
nonlinear defects create an ideal environment for the observation
of many of the nonlinear effects earlier predicted and studied in
other branches of physics. In particular, the existence of
nonlinear localized modes with the frequencies in the photonic
band gaps has already been predicted and demonstrated numerically
for several models of photonic crystals with the Kerr-type
nonlinearity
\cite{john,Mingaleev:2000-5777:PRE,Mingaleev:2001-5474:PRL}.

In this paper, we study the resonant light transmission and
localization in the photonic crystal waveguides and bends with
embedded nonlinear defects. For simplicity, we consider the case
of a photonic crystal created by a square lattice of infinite
dielectric rods, with waveguides made by removing some of the
rods. Nonlinear properties of the waveguides are controlled by
embedding the nonlinear defect rods. We demonstrate that the
effective interaction in such waveguiding structures is nonlocal,
and we suggest a novel theoretical approach, based on the
effective discrete equations, for describing both linear and
nonlinear properties of such photonic-crystal waveguides and
circuits, including the localized states at the waveguide bends.
Additionally, we study the transmission of waveguide bends and
emphasize the role of evanescent modes for the correct analysis of
their properties.

The paper is organized as follows. In Sec. 2, we introduce our
model of a two-dimensional (2D) photonic crystal and provide a
brief derivation of the effective discrete equations for the
photonic-crystal waveguides created by removed or embedded rods,
based on the Green function technique. In Sec. 3,  we apply
these discrete equations for the analysis of the transmission of
straight waveguides, including the resonant transmission through
an array of nonlinear defect rods embedded into a straight
waveguide. Additionally, we discuss a link between our approach
and the results obtained in the framework of the familiar
tight-binding approximation often used in the solid-state physics
models, and emphasize the important role of the evanescent modes
which can not be accounted for in the framework of the
tight-binding approximation. Section 4 is devoted to the study of
waveguide bends. First, we discuss the localized modes supported
by a waveguide bend, and then we analyze the transmission of the
waveguide bends. We demonstrate that the waveguide bends with
embedded nonlinear defects can be used for a very effective
control of light transmission. Section 5 concludes the paper with
a summary of the results and further applications of our approach.

\section{Effective discrete equations}

In this Section, we suggest and describe a novel theoretical
approach, based on the effective discrete equations, for
describing many of the properties of the photonic-crystal
waveguides and circuits, including the transmission spectra of
sharp waveguide bends. This is {\em an important part of our
analysis} because the properties of the photonic crystals and
photonic crystal waveguides are usually studied by solving
Maxwell's equations numerically, and such calculations are,
generally speaking, time consuming. Moreover, the numerical
solutions do not always provide a good physical insight. The
effective discrete equations we derive below and employ further in
the paper are somewhat analogous to the Kirchhoff equations for
the electric circuits. However, in contrast to electronics, in
photonic crystals both diffraction and interference become
important, and thus the resulting equations involve the long-range
interaction effects.

\begin{figure}[t]
\centerline{\hbox{
\psfig{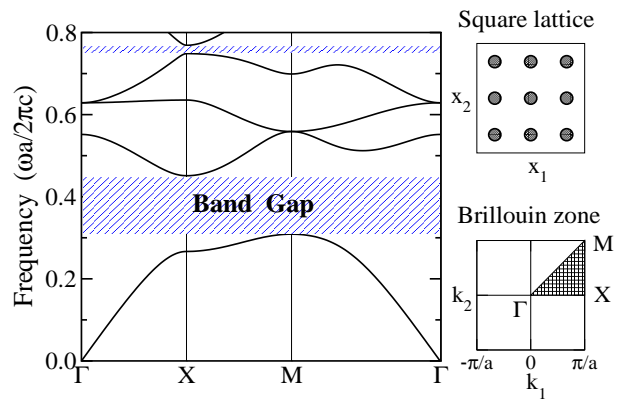}}}
\vspace{3mm} \caption{The band-gap structure of the photonic
crystal created by a square lattice of dielectric rods with
$r_0=0.18a$ and $\varepsilon_0=11.56$; the band gaps are hatched.
The top right inset shows a cross-sectional view of the 2D
photonic crystal. The bottom right inset shows the corresponding
Brillouin zone, with the irreducible zone shaded.}
\label{fig:band-r0.18}
\end{figure}

We introduce our approach for a simple model of 2D photonic
crystals consisting of infinitely long dielectric rods arranged in
the form of a square lattice with the lattice spacing $a$. We
study the light propagation in the plane normal to the rods,
assuming that the rods have a radius $r_0=0.18a$ and the
dielectric constant $\varepsilon_0=11.56$ (this corresponds to
GaAs or Si at the wavelength $\sim 1.55$ $\mu m$). For the
electric field $E(\vec{x}, t) = e^{- i \omega t} \, E(\vec{x}
\,|\, \omega)$ polarized parallel to the rods, Maxwell's equations
reduce to the eigenvalue problem
\begin{equation}
\left[ \nabla^2 + \left( \frac{\omega}{c} \right)^2
\varepsilon(\vec{x}) \right] E(\vec{x} \,|\, \omega) = 0 \; ,
\label{eq-E-omega}
\end{equation}
which can be solved by the plane-wave method
\cite{Johnson:2001-173:OE}. A perfect photonic crystal of this
type possesses a large (38\%) complete band gap between
$\omega=0.303 \times 2\pi c/a$ and $\omega=0.444 \times 2\pi c/a$
(see Fig.~\ref{fig:band-r0.18}), and it has been extensively
employed during last few years for the study of bound states
\cite{Mekis:1998-4809:PRB}, transmission of light through sharp
bends \cite{Mekis:1996-3787:PRL,Lin:1998-274:SCI}, waveguide
branches \cite{fan} and intersections \cite{johnson},
channel drop filters \cite{Fan:1998-960:PRL},
nonlinear localized modes in straight waveguides
\cite{Mingaleev:2000-5777:PRE} and discrete spatial solitons in
perfect 2D photonic crystals \cite{Mingaleev:2001-5474:PRL}.
Recently, this type of photonic crystal with a $90^o$ bent
waveguide has been fabricated in macro-porous silicon with
$a=0.57$ $\mu m$ and a complete band gap at $1.55$ $\mu m$
\cite{Zijlstra:1999-2734:JVB}.

To create a waveguide circuit, we introduce a system of defects
and assume, for simplicity, that the defects are identical rods of
the radius $r_d$ located at the points $\vec{x}_n$, where $n$ is
the index number of the defect rods. Importantly, the similar
approach can be employed equally well for the study of the defects
created by removing isolated rods in a perfect 2D lattice, and we
demonstrate such examples below.

In the photonic crystal with defects, the dielectric constant
$\varepsilon(\vec{x})$ can be presented as a sum of the periodic
and defect-induced terms, i.e.
$\varepsilon(\vec{x})=\varepsilon_{p}(\vec{x})+ \delta
\varepsilon(\vec{x})$, with
\begin{equation}
\delta \varepsilon(\vec{x}) = \sum_n \varepsilon_d [E(\vec{x}
\,|\, \omega)] \, f(\vec{x}-\vec{x}_n) \; , \label{delta-eps}
\end{equation}
where $f(\vec{x})=1$ for $|\vec{x}|<r_d$,
and it vanishes otherwise. Equation (\ref{eq-E-omega}) can be
therefore written in the following integral form
\begin{equation}
E(\vec{x} \,|\, \omega) = \left( \frac{\omega}{c} \right)^2
\int d^2\vec{y} \,\,\, G(\vec{x},
\vec{y} \,|\, \omega) \, \delta \varepsilon(\vec{y}) \,
E(\vec{y} \,|\, \omega) \; ,
\label{eq-green-int}
\end{equation}
where $G(\vec{x}, \vec{y} \,|\, \omega)$ is the Green function 
of a perfect 2D photonic crystal 
(see, e.g., Ref. \onlinecite{Mingaleev:2000-5777:PRE}).

A single defect rod is described by the function $\delta
\varepsilon(\vec{x}) = \varepsilon_d f(\vec{x})$, and it can
support one or more localized modes. Such localized modes are the
eigenmodes of the discrete spectrum of the following eigenvalue
problem,
\begin{equation}
{\cal E}_l(\vec{x}) = \left( \frac{\omega_l}{c} \right)^2
\int_{r_d} d^2\vec{y} \,\,\, G(\vec{x},
\vec{y} \,|\, \omega_l) \, \varepsilon_d f(\vec{y}) \,
{\cal E}_l(\vec{y}) \; ,
\label{eq-green-single}
\end{equation}
where $\omega_l$ is the frequency (a discrete eigenvalue) of the
$l$-th eigenmode and ${\cal E}_l(\vec{x})$ is the corresponding
electric field.

When we increase the number of defect rods (for example, in order
to create photonic-crystal waveguide circuits
\cite{Mekis:1996-3787:PRL,Lin:1998-274:SCI,fan,johnson,Fan:1998-960:PRL}),
the numerical solution of the integral equation
(\ref{eq-green-int}) becomes complicated and, moreover, it is
severely restricted by the current computer facilities. Therefore,
one of our major goals in this paper is to develop a new
approximate physical model that would allow the application of
fast numerical techniques, combined with a reasonably good
accuracy, for the study of more complicated (linear and nonlinear)
waveguide circuits in photonic crystals.

To achieve our goal, we consider the localized states created by a
(in general, complex) system of defects (\ref{delta-eps}) as a
linear combination of the localized modes ${\cal E}_l(\vec{x})$
supported by isolated defects:
\begin{equation}
E(\vec{x} \,|\, \omega) = \sum_{l,n} \psi_n^{(l)} (\omega)
{\cal E}_l(\vec{x}-\vec{x}_n) \; .
\label{E-TB}
\end{equation}
Substituting Eq.~(\ref{E-TB}) into Eq.~(\ref{eq-green-int}),
multiplying it by ${\cal E}_{l'}(\vec{x}-\vec{x}_{n'})$ and
integrating with $\vec{x}$, we obtain a system of discrete
equations for the amplitudes $\psi_n^{(l)}$ of the $l$-th
eigenmodes localized at $n$-th defect rods: 
\begin{equation}
\sum_{l,n} \lambda_{l,n}^{l',n'} \psi_n^{(l)} =
\sum_{l,n,m} \varepsilon_d \mu_{l,n,m}^{l',n'}(\omega)
\psi_n^{(l)} \; ,
\label{eq-discrete}
\end{equation}
where
\begin{eqnarray}
&&\lambda_{l,n}^{l',n'} = \int d^2\vec{x} \,\,\,
{\cal E}_l(\vec{x}-\vec{x}_n) \,
{\cal E}_{l'}(\vec{x}-\vec{x}_{n'}) \; , \nonumber \\ \nonumber
&&\mu_{l,n,m}^{l',n'}(\omega) = \left( \frac{\omega}{c} \right)^2
\int d^2\vec{x} \,\, {\cal E}_{l'}(\vec{x}-\vec{x}_{n'}) \times \\
&&~~~~~~~~~\int d^2\vec{y} \,\,\, G(\vec{x},
\vec{y} \,|\, \omega) \, f(\vec{y}-\vec{x}_m)
{\cal E}_l(\vec{y}-\vec{x}_n) \; .
\label{coeffs}
\end{eqnarray}
It should be emphasized that the discrete equations
(\ref{eq-discrete})--(\ref{coeffs}) are derived by using only the
approximation provided by the ansatz (\ref{E-TB}). As can be
demonstrated by comparing the approximate results with the direct
numerical solutions of the Maxwell equations, this approximation
is usually very accurate,  and it can be used in many physical
problems.

However, the effective discrete equations (\ref{eq-discrete}) are
still too complicated and, in some cases, they can be simplified
further still remaining very accurate. A good example is the case
of the photonic crystal waveguides created by a sequence of
largely separated defect rods. Such waveguides are known as {\em
the coupled-resonator optical waveguides} (CROWs)
\cite{Yariv:OL,Yariv:JOSB} or {\em coupled-cavity waveguides}
(CCWs) \cite{Bayindir:PRB}. For those cases, the localized modes
are located at each of the defect sites being only weakly coupled
with the similar neighboring modes. As is known, such a situation
can be described  very accurately by the so-called {\em
tight-binding approximation} (see also Ref.~\onlinecite{tb}). 
For our formalism, this means that 
$\lambda_{l,n}^{l',n'}=\mu_{l,n,m}^{l',n'}=0$ 
for $|n'-n|>1$ and $|n'-m|>1$. The most
important feature of the CROW circuits is that their bends are
reflectionless throughout the entire band
\cite{Yariv:OL,Bayindir:PRB}. This is in a sharp contrast with the
conventional photonic crystal waveguides created by a sequence of
the removed or introduced defect rods (see e.g., 
Ref.~\onlinecite{Mekis:1996-3787:PRL} and references therein) 
where the 100\%
transmission through a waveguide bend is known to occur only at
certain resonant frequencies. In spite of this visible advantage,
the CROW structures have a very narrow guiding band and, as a
result, effectively they demonstrate a complete transmission
through the waveguide bend in a narrow frequency interval too.

Below, we consider different types of photonic crystal waveguides
and show that a very accurate simplification of
Eqs.~(\ref{eq-discrete}) is provided by accounting for {\em an
indirect coupling} between the remote defect modes, caused by the
slowly decaying Green function, $\mu_{l,n,m}^{l',n'} \neq 0$ for
$|n'-n| \leq L$, where the number $L$ of effectively coupled
defects lies usually between five and ten. As we show below, this
type of interaction, which is neglected in the tight-binding
approximation, is important for understanding the transmission
properties of the photonic-crystal waveguides. At the same time,
we neglect a direct overlap between the nearest-neighbor
eigenmodes, which is often considered to be important
\cite{Yariv:OL,Bayindir:PRB}), i.e. we consider
$\lambda_{l,n}^{l',n'}=\delta_{l,l'} \, \delta_{n,n'}$ (with
$\delta_{l, \, l'}$ being the Dirac delta function) and
$\mu_{l,n,m}^{l',n'}=0$ for $n \neq m$. Taking into account this
interaction leads to negligible corrections only.

Assuming that the defects support only {\em the monopole
eigenmodes} (marked by $l=1$), the coefficients (\ref{coeffs})
can be calculated reasonably accurately in approximation
that the electric field remains constant inside
the defect rods, i.e.
${\cal E}_1(\vec{x}) \sim f(\vec{x})$.
This corresponds to the averaging of the electric field
in the integral equation (\ref{eq-green-int}) over the
cross-section of the defect rods
\cite{McGurn:1996-7059:PRB,Mingaleev:2000-5777:PRE}.
In this case the resulting
approximate discrete equation for the amplitudes of the
electric fields $E_n(\omega) \equiv \psi_n^{1}(\omega)$
of the eigenmodes excited at
the defect sites has the following matrix form:
\begin{eqnarray}
\sum_m M_{n,m}(\omega) E_m(\omega) = 0 \; , \nonumber \\
M_{n,m}(\omega) = \varepsilon_{d}(E_m) \,
J_{n,m}(\omega) - \delta_{n,m} \; ,
\label{eq-E-disc}
\end{eqnarray}
where $J_{n,m}(\omega) \equiv \mu_{1,m,m}^{1,n}(\omega)$
is a coupling constant calculated in the approximation that
${\cal E}_1(\vec{x}) \sim f(\vec{x})$, so that
\begin{equation}
J_{n,m}(\omega) = \left( \frac{\omega}{c} \right)^2
\int_{r_d} d^2 \vec{y} \,\,\,
G(\vec{x}_n, \vec{x}_m + \vec{y} \,|\, \omega )
\label{Jnm}
\end{equation}
is completely determined by the Green function of a
perfect 2D photonic crystal (see details in 
Refs.~\onlinecite{Mingaleev:2000-5777:PRE,Mingaleev:2001-5474:PRL}).

First of all, we check the accuracy of our approximate model
(\ref{eq-E-disc}) for the case of {\em a single defect} located at
the point $\vec{x}_0$. In this case, Eq.~(\ref{eq-E-disc}) yields
the simple result $J_{0,0}(\omega_{d})= 1/\varepsilon_{d}$, and
this expression defines the frequency $\omega_{d}$ of the defect
mode. In particular, applying this result to the case when the
defect is created by a single removed rod, we obtain the frequency
$\omega_{d}=0.391 \times 2\pi c/a$ which differs only by $1\%$
from the value $\omega_{d}=0.387 \times 2\pi c/a$ calculated with
the help of the MIT Photonic-Bands numerical code
\cite{Johnson:2001-173:OE}.

\section{Straight waveguides}

\subsection{Waveguide dispersion}

\begin{figure}
\vspace*{0mm} \centerline{\hbox{
\psfig{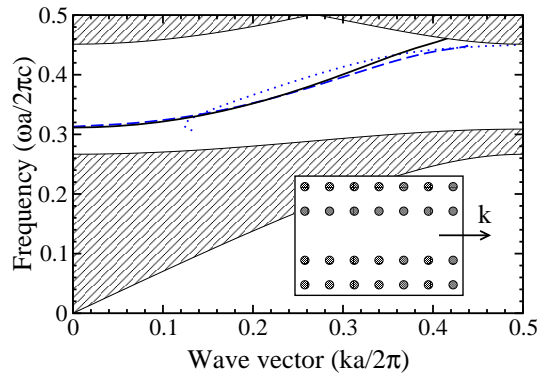}}}
\vspace*{2mm} \caption{Dispersion relation for a 2D
photonic-crystal waveguide (shown in the inset) calculated by the
super-cell method \protect\cite{Johnson:2001-173:OE} (solid), and
from the approximate equations
(\ref{transfer-matrix})--(\ref{transfer-vect}) for $L$=7 (dashed)
and $L$=1 (dotted). The hatched areas are the projected band
structure of a perfect 2D crystal.} \label{fig:dispersion}
\end{figure}

A simple single-mode waveguide can be created by removing a row of
rods (see the inset in Fig. \ref{fig:dispersion}). Assuming that
the waveguide is straight ($M_{n,m} \equiv M_{n-m}$) and
neglecting the coupling between the remote defect rods (i.e.
$M_{n-m}=0$ for all $|n-m|>L$), we rewrite Eq. (\ref{eq-E-disc})
in the transfer-matrix form, $\vec{F}_{n+1}=\hat{T} \vec{F}_{n}$,
where we introduce the vector $\vec{F}_{n} = \{ \, E_n, \,
E_{n-1}, \, ... \, , \, E_{n-2L+1} \, \}$ and the transfer matrix
$\hat{T} = \{ T_{i,j} \}$ with the non-zero elements
\begin{eqnarray}
&& T_{1,j}(\omega)= -\frac{M_{L-j}(\omega)}{M_{L}(\omega)}
\quad \mbox{for} \quad j=1, 2, ... , 2L \; , \nonumber \\
&& T_{j,j+1}=1 \quad \mbox{for} \quad j=1, 2, ... , 2L-1 \; .
\label{transfer-matrix}
\end{eqnarray}
Solving the eigenvalue problem
\begin{eqnarray}
\hat{T}(\omega) \vec{\Phi}^{p} = \exp\{i k_p(\omega)\} \,
\vec{\Phi}^{p} \; ,
\label{transfer-vect}
\end{eqnarray}
we can find the $2L$ eigenmodes of the photonic-crystal waveguide.
The eigenmodes with real wavenumbers $k_p(\omega)$ correspond to
the modes propagating along the waveguide. In the waveguide shown
in Fig. \ref{fig:dispersion}, there exist only two such modes (we
denote them as $\vec{\Phi}^{1}$ and $\vec{\Phi}^{2}$), propagating
in the opposite directions ($k_1=-k_2>0$). In
Fig.~\ref{fig:dispersion} we plot the dispersion relation
$k_1(\omega)$  calculated by {\em three different methods}: first
(solid curve) is calculated directly by the super-cell method
\cite{Johnson:2001-173:OE}, and other two are found from
Eq.~(\ref{transfer-vect}) in the nearest-neighbor approximation
($L$=1, dotted curve) corresponding to the tight-binding models,
and also by taking into account the long-range interaction through
the coupling between several neighbors ($L$=7, dashed curve).
Indeed, we observe a very good agreement for the conditions when
the long-range interaction is taken into account.

\begin{figure}
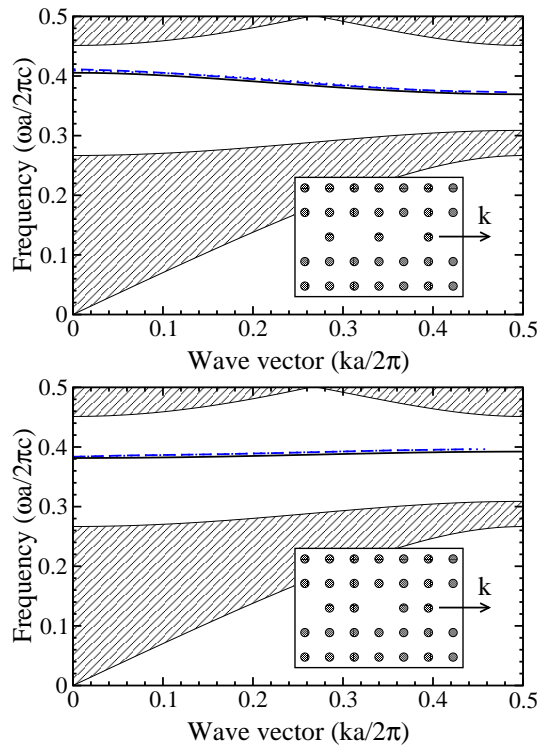

\vspace*{0mm} \centerline{\hbox{
\psfig{figure=josb-pbg-3a.eps,clip=,width=70mm,angle=0}}}
\centerline{\hbox{
\psfig{figure=josb-pbg-3b.eps,clip=,width=70mm,angle=0}}}
\vspace*{2mm} \caption{The same as in Fig. \ref{fig:dispersion}
but for two other types of waveguides better described by the
tight-binding models.} \label{fig:dispersion1}
\end{figure}

It is important to compare the results provided by our method with
those obtained in the tight-binding approximation. For the
waveguide shown in  Fig. \ref{fig:dispersion}, such an
approximation, rigorously speaking, is not valid, but its analog
can be considered at least formally for the case of the
nearest-neighbor interaction when in Eqs.
(\ref{transfer-matrix})--(\ref{transfer-vect}) we take $L$=1.  The
interaction between the remote rods cannot be neglected as soon as
we study the waveguides created by removing (or inserting) all
rods along the row or more complicated structures of this type. In
such a case, as is seen from Fig.~\ref{fig:dispersion}, the
dispersion relation found in the tight-binding approximation is
incorrect and, in order to obtain accurate results, one should
take into account the coupling between several defect rods. We
verify that this statement is also valid for multi-mode
waveguides, e.g. those created by removing several rows of rods.

\begin{figure}
\vspace*{0mm} \centerline{\hbox{
\psfig{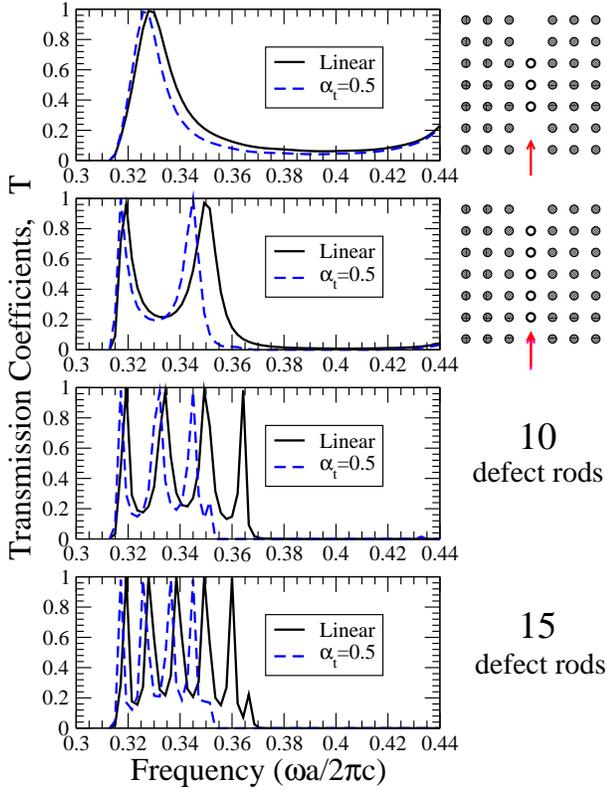}}}
\vspace*{2mm} \caption{Transmission coefficients of an array of
nonlinear defect rods calculated from
Eqs.~(\ref{eq-E-disc})--(\ref{waves-out}) with $L=7$ in the linear
limit of very small $|\alpha_t|^2$ (solid curves) and for the
nonlinear transmission when the output intensity is
$|\alpha_t|^2=0.25$ (dashed curves), for different numbers of the
defects. We use nonlinear defect rods with the dielectric constant
$\varepsilon_d^{(0)}=7$; they are marked by open circles on the
right insets.} 
\label{fig:scatt-nonlin}
\end{figure}

However, for the waveguides of a different geometry, when only the
second (or third, etc) rods are removed, all methods provide a
reasonably good agreement with the direct numerical results, as is
shown for two examples in Fig. \ref{fig:dispersion1}. In this
case, the waveguides are created by an array of cavity modes,  and
they are similar to the CROW structures earlier analyzed by
several authors \cite{Yariv:OL,Yariv:JOSB,Bayindir:PRB}. Thus, the
dispersion properties of CROWs (or similar waveguides) can be
described with a good accuracy by the tight-binding approximation;
our new approach confirms this conclusion, and it provides a
simple method for the derivation of the approximate equations.

\subsection{Resonant transmission of an array of defects}
\label{sec:transmission}

In addition to the propagating guided modes, in photonic crystal
waveguides there always exist {\em evanescent modes} with
imaginary $k_p$. These modes cannot be accounted for in the
framework of the tight-binding theory that relies on the
nearest-neighbor interaction between the defect rods. Although the
evanescent modes remain somewhat ``hidden'' in straight
waveguides, they become crucially important in more elaborated
structures such as waveguides with embedded linear or nonlinear
defects, waveguide bends and branches. In these cases the
evanescent modes manifest themselves in several different ways. In
particular, they determine {\em non-trivial transmission
properties} of the photonic-crystal circuits considered below.

As the first example of the application of our approach, we study
the transmission of a straight waveguide with embedded nonlinear
defects. Such a structure can be considered as two semi-infinite
straight waveguides coupled by a finite region of defects located
between them. The coupling region may include both linear (as a
domain of a perfect waveguide) and nonlinear (embedded) defects.
We assume that the defect rods inside the coupling region are
characterized by the index that runs from $a$ to $b$, and the
amplitudes $E_m$ ($m=a, \ldots, b$) of the electric field at the
defects are all unknown. We number the guided modes
(\ref{transfer-vect}) in the following way: $p=1$ corresponds to
the mode propagating in the direction of the nonlinear section
(for both ends of the waveguide), $p=2$ corresponds to the mode,
propagating in the opposite direction, $p=3, ... , L+1$ correspond
to the evanescent modes which grow in the direction of the
nonlinear section, and $p=L+2, ... , 2L$ correspond to the
evanescent modes which decay in the direction of the nonlinear
section. Then, we can write the incoming and outgoing waves in the
semi-infinite waveguide sections as a superposition of the guided
modes:
\begin{eqnarray}
E^{in}_m &=& \alpha_i \Phi^1_{a-m} + \alpha_r \Phi^2_{a-m} +
\sum_{p=3}^{L+1} \beta^{in}_p \Phi^p_{a-m} \; ,
\label{waves-in}
\end{eqnarray}
for $m=a-2L, ... , a-1$, and
\begin{eqnarray}
E^{out}_m &=& \alpha_t \Phi^2_{m-b} +
\sum_{p=3}^{L+1} \beta^{out}_p \Phi^p_{m-b} \; ,
\label{waves-out}
\end{eqnarray}
for $m=b+1, ... , b+2L$, where $\beta^{in}_p$ and $\beta^{out}_p$
are unknown amplitudes of the evanescent modes growing in the
direction of the nonlinear section,  whereas $\alpha_i$,
$\alpha_t$ and $\alpha_r$ are unknown amplitudes of the incoming,
transmitted, and reflected propagating waves. We take into account
that the evanescent modes with $p>L+1$ (growing in the direction
of waveguide ends) must vanish. 
Now, substituting Eqs.~(\ref{waves-in})--(\ref{waves-out}) 
into Eq.~(\ref{eq-E-disc}), we
obtain a system of linear (or nonlinear, for nonlinear defects) 
equations with $2L+b-a+1$ unknown.
Solving this system, we find the transmission coefficient, $T=
|\alpha_t/\alpha_i|^2$, and reflection coefficient,
$R=|\alpha_r/\alpha_i|^2$, as functions of the light frequency,
$\omega$, and intensity, $|\alpha_i|^2$ or $|\alpha_t|^2$.
Recently, we have demonstrated \cite{MK:2002:OL} that the linear
transmission properties of the waveguide bends are described very
accurately by this approach. Below, we study {\em nonlinear
transmission} of the photonic-crystal waveguides and waveguide
bends.

\begin{figure}
\vspace*{0mm} \centerline{\hbox{
\psfig{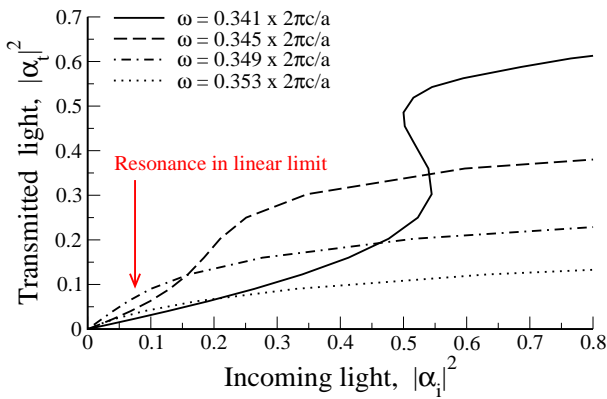}}}
\vspace*{2mm} \caption{Bistability in the nonlinear transmission
of an array of five nonlinear defect rods shown in
Fig.~\ref{fig:scatt-nonlin}(b).}
\label{fig:scatt}
\end{figure}

In Fig.~\ref{fig:scatt-nonlin}, we present our results for the
transmission spectra of the straight waveguides (created by a row
of removed rods) with an array of embedded nonlinear defects. 
We assume throughout the paper that all nonlinear defect rods 
are identical, with the radius $r_d=r_0=0.18a$ and the dielectric 
constant, $\varepsilon_d=\varepsilon_d^{(0)}+|E_n|^2$ 
(with~~$\varepsilon_d^{(0)}=7$), which grows 
linearly with the light intensity (the so-called Kerr effect). 
In the linear limit,  the embedded defects behave like an
effective resonant filter, and only the waves with some specific
resonance frequencies can be effectively transmitted through the
defect section. The resonances appear due to the excitation of
cavity modes inside the defect region, whereas a single defect
does not demonstrate any resonant behavior. When the 
intensity of the input wave grows, the resonant frequencies 
found in the linear limit get shifted to lower values. 
The sensitivity of different resonances to the change of 
the light intensity is quite different and may be tuned by 
matching of the defect parameters. 
The nonlinear resonant transmission
is found to possess {\em bistability}, similar to another problem of
the nonlinear transmission (see, e.g., Refs.~\onlinecite{bistab}). 
The bistable transmission occurs for the frequencies smaller then
the resonant, in a linear limit, frequency (see Fig. \ref{fig:scatt}).

\subsection{An optical diode}

An all-optical ``diode'' is a spatially nonreciprocal device which
allows unidirectional propagation of a signal at a given
wavelength. In the ideal case, the diode transmission is 100\% in
the ``forward'' propagation, while it is much smaller or vanishes
for ``backward'' (opposite) propagation, yielding a unitary
contrast.

The first study of the operational mechanism for a passive optical
diode based on a photonic band gap material was carried out by
Scalora {\em et al.} \cite{diode,diode1}. These authors considered
the pulse propagation near the band edge of a one-dimensional
photonic crystal structure with a spatial gradiation in the linear
refractive index, together with a nonlinear medium response, and
found that such a structure can result in unidirectional pulse
propagation.

\begin{figure}
\vspace*{0mm} \centerline{\hbox{
\psfig{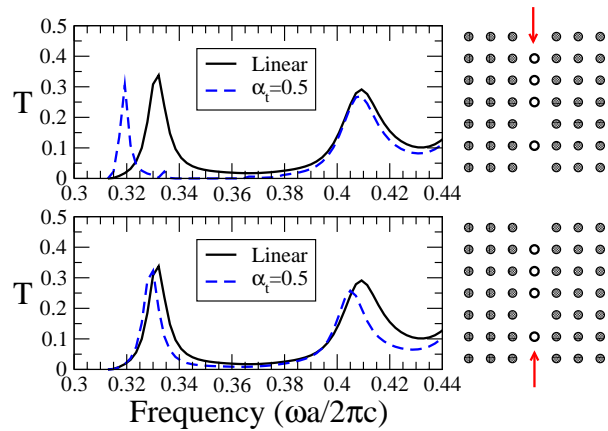}}}
\vspace*{2mm} \caption{Transmission coefficients of an 
asymmetric array of nonlinear defect rods calculated for 
the same parameters as in Fig.~\ref{fig:scatt-nonlin}.} 
\label{fig:scatt1}
\end{figure}

\begin{figure}
\vspace*{0mm} \centerline{\hbox{
\psfig{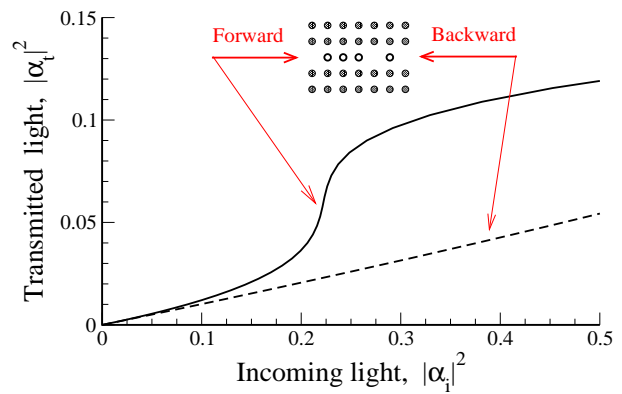}}}
\vspace*{2mm} \caption{Nonlinear transmission of the
optical diode for the forward (see top of Fig.~\ref{fig:scatt1})
and backward (see bottom of Fig.~\ref{fig:scatt1})
directions at the light frequency $\omega=0.326 \times 2\pi c/a$.}
\label{fig:scatt2}
\end{figure}

To implement this concept for the waveguide geometry discussed
above, we consider the asymmetric structure made of four nonlinear
defect rods, as shown in the right insets on Fig.
\ref{fig:scatt1}. Figure \ref{fig:scatt1} shows the transmission
spectra of such an asymmetric structure in the opposite directions
indicated by two arrows in the right insets. As is seen, 
in the linear limit the transmission is characterized by two resonant 
frequencies and does not depend on the propagation direction. 
However, since the sensitivity of both resonant frequencies 
to the change of the light intensity is different for the ``forward'' 
(see Fig.~\ref{fig:scatt1}, top) and ``backward'' (see
Fig.~\ref{fig:scatt1}, bottom) propagation directions, 
the transmission becomes, in the vicinity of resonant frequencies, 
highly asymmetric for large input intensities.
This results into nearly unidirectional waveguide transmission, as
shown in Fig.~\ref{fig:scatt2}. 

In contrast to the perfect resonators
used in Fig.~\ref{fig:scatt-nonlin}, the transmission of the
asymmetric structure under consideration is not very efficient 
at the resonant frequencies. 
However, we believe
that the optical diode effect, with much better efficiency, can be
found in other types of the waveguide geometry and a unitary
contrast can be achieved by a proper optimization of the waveguide
and defect parameters, that can be carried out by employing our
method and the effective discrete equations derived above.

\section{Waveguide bends}

\begin{figure}
\vspace*{0mm} \centerline{\hbox{
\psfig{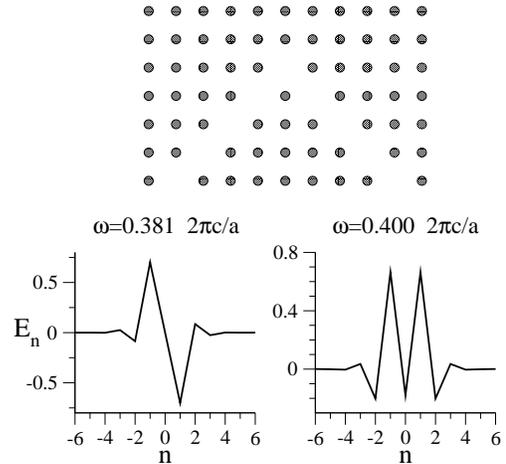}}}
\vspace*{2mm} \caption{Electric field $E_n$ for two bound states
supported by a $90^o$ waveguide bend shown in the top. Center of
the bend is located at $n=0$.} \label{fig:bound}
\end{figure}

In addition to the non-trivial transmission properties, the
evanescent modes manifest themselves in creation of {\em localized
bound states} in the vicinity of the waveguide bends or branches.
To demonstrate this effect, we consider the simplest case of a
sharp bend of the waveguide created by a row of removed rods. As
was shown in Ref.~\onlinecite{Mekis:1998-4809:PRB}, in the cases when
the waveguide bend can be considered as a finite section of a
waveguide of different type, the bound states correspond closely
to cavity modes excited in this finite section. However, such a
simplified one-dimensional model does not describe correctly more
complicated cases \cite{Mekis:1998-4809:PRB}, 
even the properties of the bent waveguide
depicted in Fig.~\ref{fig:bound}. The
situation becomes even more complicated for the waveguide branches
\cite{fan}. In contrast, solving the discrete equations
(\ref{eq-E-disc}) we can find the frequencies and profiles of the
bound states excited in an arbitrary complex set of defects. For
particular case of the waveguide bend shown in
Fig.~\ref{fig:bound}, we find two bound states localized at the
bend and their profiles (cf. our Fig.~\ref{fig:bound} with Fig.~9 in
Ref.~\onlinecite{Mekis:1998-4809:PRB}). It should be noted that the
frequencies of the modes are found from Eq. (\ref{eq-E-disc}) with
the accuracy of $1.5\%$.

Additionally, the evanescent modes determine the non-trivial
transmission properties of the waveguide bends which can also be
calculated with the help of our discrete equations. To demonstrate
these features, we consider a bent waveguide consisting of two
coupled semi-infinite straight waveguides with a finite section of
defects between them. The finite section includes a bend with a
safety margin of the straight waveguide at both ends.

\begin{figure}
\vspace*{0mm} \centerline{\hbox{
\psfig{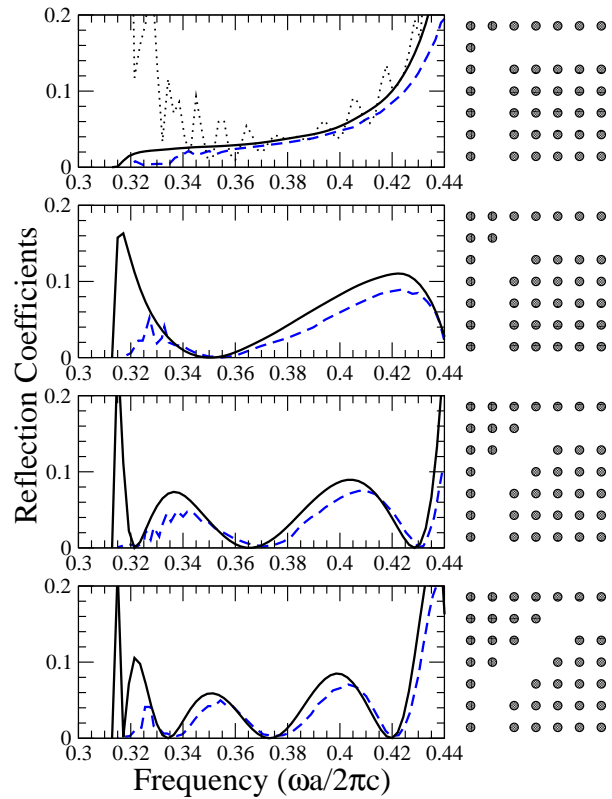}}}
\vspace*{2mm} \caption{Reflection coefficients calculated by the
finite-difference time-domain method (dashed, from 
Ref.~\protect\onlinecite{Mekis:1996-3787:PRL}) and from 
Eqs.~(\ref{eq-E-disc})--(\ref{waves-out}) with $L=7$ (full lines)
and $L=2$ (dotted, only in the top plot), for different bend
geometries.} \label{fig:scatt-bend}
\end{figure}

Similar to what we discussed in Sec.~\ref{sec:transmission} 
for the straight waveguides, we
solve the system of effective discrete equations to find the
transmission, $T=|\alpha_t/\alpha_i|^2$, and reflection,
$R=|\alpha_r/\alpha_i|^2$, coefficients of the waveguide bends. In
Fig.~\ref{fig:scatt-bend} we present our results for the
transmission spectra of several types of bent waveguides, which
have been discussed in Ref.~\onlinecite{Mekis:1996-3787:PRL}, where 
the possibility of high transmission through sharp bends in
photonic-crystal waveguides was first demonstrated. We compare the
reflection coefficients calculated by the finite-difference
time-domain method in Ref.~\onlinecite{Mekis:1996-3787:PRL} 
(dashed lines) with our results, calculated from
Eqs.~(\ref{eq-E-disc})--(\ref{waves-out}) for $L=7$ (full
lines) and for $L=2$ (dotted line in the top plot). As is clearly seen,
Eqs.~(\ref{eq-E-disc})--(\ref{waves-out}) provide a very accurate
method for calculating the transmission spectra of the waveguide
bends, if only we account for long-range interactions. It should
be emphasized that the approximation, in which the only next-neighbor
interactions are taken in to account, is usually too crude, while
the tight-binding theory incorrectly predicts a perfect transmission 
for all guiding frequencies.

\begin{figure}
\vspace*{0mm} \centerline{\hbox{
\psfig{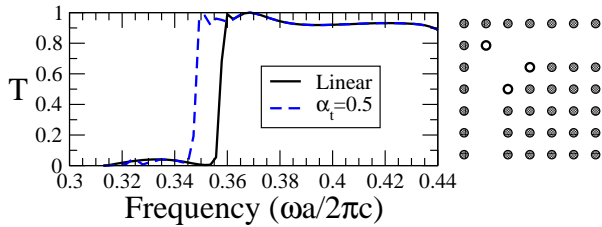}}}
\vspace*{2mm} \caption{Transmission of a waveguide bend with three
embedded nonlinear defect rods in the linear (solid curve) and
nonlinear (dashed curve) regimes. Defect rods have the dielectric
constant $\varepsilon_d^{(0)}=7$, and they are marked by open
circles.} \label{fig:scatt-bend-nl}
\end{figure}

\begin{figure}
\vspace*{0mm} \centerline{\hbox{
\psfig{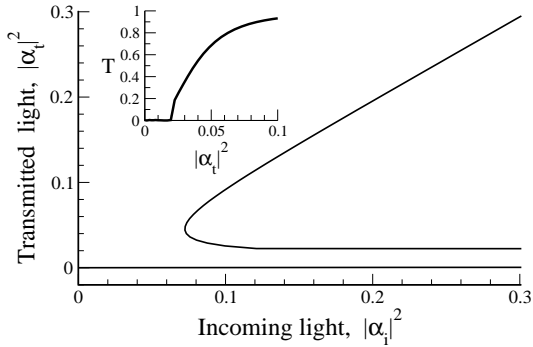}}}
\vspace*{2mm} \caption{Bistable nonlinear transmission through the
waveguide bend shown in Fig.~\ref{fig:scatt-bend-nl}, for the
light frequency $\omega=0.351 \times 2\pi c/a$.}
\label{fig:nonlin-bend-bistab}
\end{figure}

The resonant transmission can be modified dramatically by
introducing both linear and/or nonlinear defects into the waveguide
bends. To illustrate such features, we consider the waveguide bend
with three embedded nonlinear defects, as is depicted in
Fig.~\ref{fig:scatt-bend-nl} on the right inset, where these defects
are shown by open circles. In the linear regime, such a sharp bend
behaves as an optical threshold device that efficiently transmits
the guided waves with frequencies above the threshold one, but
completely reflects the waves with the lower frequencies. The
transmission coefficient of this waveguide bend in the linear
limit is shown in Fig.~\ref{fig:scatt-bend-nl} by a solid curve.
When the input intensity increases, the threshold frequency
decreases, extending the transmission region (see a dashed curve
in Fig.~\ref{fig:scatt-bend-nl}). The resulting transmission as a
function of the input intensity demonstrates a sharp nonlinear
threshold character with an extremely low transmission of the
waves below a certain (rather small) threshold intensity, see 
Fig.~\ref{fig:nonlin-bend-bistab}.

\section{Conclusions}

We have suggested a novel theoretical approach for describing a
broad range of transmission properties of the photonic crystal
waveguides and circuits. Our approach is based on the analysis of
the effective discrete equations derived with the help of the
Green function technique, and it generalizes the familiar
tight-binding approximations usually employed to study the
coupled-resonator or coupled-cavity optical waveguides. The
effective discrete equations we have introduced in this paper
emphasize the important role played by the evanescent modes in the
transmission characteristics of the photonic crystal circuits with
waveguide bends and/or embedded defects. Employing this technique,
we have studied the properties of several important elements of
the (linear and nonlinear) photonic crystal circuits, including a
nonlinear bistable transmitter and an optical diode created by an
asymmetric structure of nonlinear defects. We believe that our
approach can be useful for solving more complicated problems, and
it can be applied to study the transmission characteristics of the
waveguide branches, channel drop filters, etc.

The work has been partially supported by the Australian Research
Council.


\end{multicols}


\begin{thebibliography}{99}
\bibitem{book} J. D. Joannoupoulos, R. B. Meade, and
J. N. Winn, {\em Photonic Crystals: Molding the Flow of Light}
(Princeton University Press, Princeton N.J., 1995).

\bibitem{book_sakoda} K. Sakoda, {\em Optical Properties of
Photonic Crystals} (Springer-Verlag, Berlin, 2001); see also a
comprehensive review paper, T.F. Krauss and R.M. De la Rue, {\em
``Photonic crystals in the optical regime--past, present and
future''}, Prog. Quantum Electron. {\bf 23}, 51-96 (1999), and
references therein.

\bibitem{john2} See, e.g., K. Busch and S. John,
{\em ``Liquid-crystal photonic-band-gap
materials: The tunable electromagnetic vacuum''}, Phys. Rev. Lett.
{\bf 83}, 967-970 (1999), and discussions therein.

\bibitem{john} S. John and N. Ak\"ozbek,
{\em ``Nonlinear optical solitary waves in
a photonic band gap''}, Phys. Rev. Lett. {\bf 71}, 1168-1171 (1993);
{\em ``Optical solitary waves in two- and three-dimensional
nonlinear photonic band-gap structures''}, Phys. Rev. E {\bf 57},
2287-2319 (1998).

\bibitem{Mingaleev:2000-5777:PRE} S.F. Mingaleev, Yu.S. Kivshar,
and R.A. Sammut,
{\em ``Long-range interaction and nonlinear localized modes in
photonic crystal waveguides''},
Phys. Rev. E {\bf 62}, 5777-5782 (2000).

\bibitem{Mingaleev:2001-5474:PRL}  S.F. Mingaleev
and Yu.S. Kivshar, {\em ``Self-trapping and stable localized modes
in nonlinear photonic crystals''}, Phys. Rev. Lett. {\bf 86},
5474-5477 (2001).

\bibitem{Johnson:2001-173:OE} S.G. Johnson and J.D. Joannopoulos,
{\em ``Block-iterative frequency-domain methods for Maxwell's
equations in a planewave basis''},
Optics Express {\bf 8}, 173-190 (2001).

\bibitem{Mekis:1998-4809:PRB} A. Mekis, S.H. Fan, and J.D.
Joannopoulos, {\em ``Bound states in photonic crystal waveguides
and waveguide bends''}, Phys. Rev. B {\bf 58}, 4809-4817 (1998).

\bibitem{Mekis:1996-3787:PRL} A. Mekis, J.C. Chen, I. Kurland,
S.H. Fan, P.R. Villeneuve, and J.D. Joannopoulos,
{\em ``High Transmission through Sharp Bends in Photonic
Crystal Waveguides''}, Phys. Rev. Lett. {\bf 77}, 3787-3790 (1996).

\bibitem{Lin:1998-274:SCI} S.Y. Lin, E. Chow, V. Hietala, P.R.
Villeneuve, and J.D. Joannopoulos,
{\em ``Experimental Demonstration of Guiding and Bending of
Electromagnetic Waves in a Photonic Crystal''},
Science {\bf 282}, 274-276 (1998).

\bibitem{fan} S. Fan, S.G. Johnson, J.D. Joannopoulos, C.
Manolatou, and H.A. Haus,
{\em ``Waveguide branches in photonic crystals''},
J. Opt. Soc. Am. B {\bf 18}, 162-165 (2001).

\bibitem{johnson}
S. G. Johnson, C. Manolatou, S. Fan, P. R. Villeneuve,
J. D. Joannopoulos, and H. A. Haus,
{\em ``Elimination of cross talk in waveguide intersections''},
Opt. Lett. {\bf 23}, 1855-1857 (1998).

\bibitem{Fan:1998-960:PRL} S.H. Fan, P.R. Villeneuve, and J.D.
Joannopoulos, {\em ``Channel drop tunneling through localized
states''}, Phys. Rev. Lett. {\bf 80}, 960-963 (1998).

\bibitem{Zijlstra:1999-2734:JVB} T. Zijlstra, E. van der Drift,
M.J.A. de Dood, E. Snoeks, and A. Polman, {\em ``Fabrication of
two-dimensional photonic crystal waveguides for 1.5 $\mu$m in
silicon by deep anisotropic dry etching''}, J. Vac. Sci. Technol.
B {\bf 17}, 2734-2739 (1999).

\bibitem{Yariv:OL} A. Yariv, Y. Xu, R.K. Lee, and A. Scherer,
{\em ``Coupled-resonator optical waveguide: a proposal and
analysis''}, Opt. Lett. {\bf 24}, 711-713 (1999).

\bibitem{Yariv:JOSB} Y. Xu, R.K. Lee, and A. Yariv, {\em ``Propagation and
second-harmonic generation of electromagnetic waves in a
coupled-resonator optical waveguide''}, J. Opt. Soc. Am. B {\bf
17}, 387-400 (2000).

\bibitem{Bayindir:PRB} M. Bayindir, B. Temelkuran, and E. Ozbay, {\em ``Propagation
of photons by hopping: A waveguiding mechanism through localized
coupled cavities in three-dimensional photonic crystals''}, Phys.
Rev. B {\bf 61}, R11855-11858 (2000).

\bibitem{tb} E. Lidorikis, M.M. Sigalas, E. Economou, and C.M.
Soukoulis, {\em ``Tight-binding parametrization for photonic band
gap materials''}, Phys. Rev. Lett. {\bf 81}, 1405-1408 (1998).


\bibitem{McGurn:1996-7059:PRB}
A.R. McGurn, {\em ``Green's-function theory for row and periodic
defect arrays in photonic band structures''}, Phys. Rev. B {\bf
53}, 7059-7064 (1996).

\bibitem{MK:2002:OL}
S.F. Mingaleev and Yu.S. Kivshar, {\em ``Effective equations for
photonic crystal waveguides and circuits''}, Opt. Lett. {\bf 27},
231-233 (2002).

\bibitem{bistab}  F. Delyon, Y.-E. L\'evy, and B. Souillard, {\em
``Nonperturbative bistability in periodic nonlinear media''},
Phys. Rev. Lett. {\bf 57}, 2010-2013 (1986);  Q. Li, C.T. Chan,
K.M. Ho, and C.M. Soukoulis, {\em ``Wave propagation in nonlinear
photonic band-gap materials''}, Phys. Rev. B {\bf 53}, 15577-15585
(1996); E. Centero and D. Felbacq, {\em ``Optical bistability in
finite-size nonlinear bidimentional photonic crystals doped by a
microcavity''}, Phys. Rev. B {\bf 62}, R7683-R7686 (2000).

\bibitem{diode} M. Scalora, J.R. Dowling, C.M. Bowden, and M.J.
Bloemer, {\em ``The photonic band edge optical diode''}, J. Appl.
Phys. {\bf 76}, 2023-2026 (1994).

\bibitem{diode1} M.D. Tocci, M.J. Bloemer, M. Scalora, J.P.
Dowling, and C.M. Bowden, {\em ``Thin-film nonlinear optical
diode''}, Appl. Phys. Lett. {\bf 66}, 2324-2326 (1995).
\end{thebibliography}
\end{document}